Commentary

# Re-visiting the intracellular pathway of transferrin on board of a mathematical simulation

**Franco Nieto[a] and Luis S. Mayorga[ab]***

[a]Instituto de Histología y Embriología de Mendoza (IHEM) - Universidad Nacional de Cuyo - CONICET, Mendoza 5500, Argentina.
[b]Facultad de Ciencias Exactas y Naturales, Universidad Nacional de Cuyo, Mendoza 5500, Argentina.

*Corresponding author: Luis S. Mayorga (lmayorga@mendoza-conicet.gob.ar).

Franco Nieto (fnieto94@hotmail.com)

Abstract

Modeling and simulation are transforming all fields of biology. Tools like AlphaFold have revolutionized structural biology, while molecular dynamics simulations provide invaluable insights into the behavior of macromolecules in solution or on membranes. In contrast, we lack effective tools to represent the dynamic behavior of the endomembrane system. Static diagrams that connect organelles with arrows are used to depict transport across space and time but fail to specify the underlying mechanisms. This static representation obscures the dynamism of intracellular traffic, freezing it in an immobilized framework. The intracellular transport of transferrin, a key process for cellular iron delivery, is among the best-characterized trafficking pathways. In this commentary, we revisit this process using a mathematical simulation of the endomembrane system. Our model reproduces many experimental observations and highlights the strong contrast between dynamic simulations and static illustrations. This work underscores the urgent need for a consensus-based minimal functional working model for the endomembrane system and emphasizes the importance of generating more quantitative experimental data — including precise measurements of organelle size, volume, and transport kinetics— practices that were more common among cell biologists in past decades.

Introduction

A multitude of cellular processes occur in dynamic organelles that rapidly change shape, composition, and localization. These processes depend on the maturation of membrane domains, which promotes directional transport, fusion with other organelles to mix membranes and contents, and fission to segregate membrane domains and cargo. However, our current understanding of intracellular transport remains largely qualitative. Processes such as cross-presentation, cholesterol metabolism, and bacterial or viral infections—all of which occur in these dynamic organelles—are often depicted as static structures connected by arrows. These representations attempt to illustrate changes in localization, composition, and time but fall short of capturing the true complexity of intracellular dynamics. A more precise and dynamic framework for describing trafficking is urgently needed.

To illustrate the impact of adopting a dynamic perspective, we revisit the transferrin-mediated delivery of iron into cells. This pathway, thoroughly characterized in the 1980s by Lodish and other researchers, remains a foundational model in cell biology [1–4]. In a seminal study, Lodish used a set of ordinary differential equations to model transferrin transport kinetics, complemented by experimental measurements of rate constants for each step in a single round of endocytosis [1]. The process consists of the following steps:

1. Fe-loaded transferrin (Tf.Fe) binds to the transferrin receptor (R), forming the R-Tf.Fe complex.
2. R-Tf.Fe is internalized.
3. In acidic endosomes, $Fe^{3+}$ is released from transferrin, generating apo-transferrin (Tf) while the receptor-transferrin complex (R-Tf) remains intact.
4. R-Tf is recycled to the plasma membrane.
5. At the neutral pH of the extracellular environment, Tf is released from R-Tf, and the receptor is free to bind a new Tf.Fe molecule.

This stepwise description, widely taught in undergraduate cell biology courses, simplifies a highly intricate process. While steps 1 and 5 involve relatively straightforward molecular interactions, the intermediate steps (2–4) represent complex cellular processes.

For example, step 2 involves the rapid internalization of R-Tf.Fe ($t_{1/2}$ = 2.8 min [1]) via clustering into clathrin-coated pits, which invaginate and pinch off as coated vesicles. Steps 3 and 4, which are not independently measurable in Lodish's model, are estimated to have a combined $t_{1/2}$ of 5 min. These steps involve vATPase-mediated acidification of the endosome, triggering a conformational change in transferrin to release two $Fe^{3+}$ ions [5]. The released $Fe^{3+}$ is then reduced to $Fe^{2+}$ by Epithelial Antigen of the Prostate (STEAP) family proteins [6] and transported into the cytosol by Divalent Metal-Ion Transporter-1 (DMT1) [7,8], all of which depend on the acidic endosomal pH. Finally, endosomes carrying R-Tf must fuse with the plasma membrane to recycle the receptor-apotransferrin complex.

The kinetic parameters for iron delivery impose stringent constraints on the fusion, fission, and maturation processes of the endosomes mediating transferrin endocytosis and recycling. Participating organelles must efficiently execute acidification, $Fe^{3+}$ reduction, and $Fe^{2+}$ translocation to ensure timely iron delivery. Although recent studies have provided apparent rate constants for many of these reactions, our understanding of transferrin trafficking remains rooted in the schematic diagram from Lodish's original paper (reproduced in Fig. 1C).

The aim of this commentary is to highlight the need for models that account for the dynamic interplay of organelles mediating molecular transport and the molecular reactions occurring both within these organelles and at the cell surface. By combining Agent-Based Models (ABM) to simulate intracellular transport with ordinary differential equations to model intra-organelle reactions, we can achieve a more realistic understanding of transferrin-mediated iron delivery. We argue that without such quantitative approaches, our understanding of processes within the dynamic endomembrane system will remain fragmented and qualitative.

## Representing a functional endomembrane system

We extended our previous agent-based model of the endomembrane system to represent a 15 × 15 µm cross-section of a cell, incorporating 3D organelle dynamics[9]. The areas of organelles involved in the secretory and endocytic pathways were adjusted based on experimentally observed characteristics of the endomembrane system [10]. Parameters such as the percentage of internalized plasma membrane and the volume incorporated per minute were also incorporated [10]. Transport processes included homotypic and heterotypic fusion, vesicular/tubular budding, and maturation, as illustrated in Fig. 1A. Snapshots of a simulation are shown in Fig. 1B.

In this simulated endomembrane system, both soluble and membrane-bound molecules moved according to their affinity for specific membrane domains. For example, during fission, molecules with a high affinity for particular membrane domains were enriched in the organelle carrying those domains. Molecules without such specific affinities were distributed proportionally to the surface area of the dividing organelle. Similarly, soluble cargoes were distributed according to the volume of the resulting organelles. Internal vesicles were treated as soluble cargoes, with the exception that their size restricted them from entering budding vesicles and tubules.

## Following a single cycle of transferrin internalization

Transferrin receptors undergo continuous cycles of endocytosis and recycling. To populate the endomembrane system in our model, we initialized it with 6,800 transferrin receptors, corresponding to 150,000 in a standard cell [1]) , and allowed the system to reach equilibrium. These receptors were incorporated into newly formed early endosomes (EE). During fusion and fission events, receptors preferentially partitioned into sorting endosomes (SE) and recycling endosomes (RE) based on membrane domain affinity. After two hours, the receptors achieved a stable distribution, with approximately 30% localized to the plasma membrane (PM), consistent with experimental observations [1]. At this point, the system mimicked a functional endomembrane network, with transferrin receptors cycling in and out of the cell, displaying kinetics and dynamics comparable to those in living cells. To simulate iron delivery via transferrin, as described by Lodish (Fig. 1C), we implemented the following steps:

1. Step 1: Association/dissociation of transferrin (Tf.Fe) with the receptor at the PM, modeled in COPASI (Table I).
2. Step 2: Internalization of the receptor into EE, as part of the ABM simulation.
3. Step 3: Intracellular transport of the receptor and transferrin as part of the ABM simulation. Changes in the cargoes inside each organelle (modeled in COPASI, Table I)
   a. Endosomal acidification (EE and LE).
   b. $Fe^{3+}$ dissociation from transferrin at acidic pH.
   c. Reduction of $Fe^{3+}$ to $Fe^{2+}$ and translocation to the cytosol at acidic pH.
   d. Tf binding/dissociation from the receptor (pH-regulated).
4. Step 4: Recycling of endosomes to the PM as part of the ABM simulation.
5. Step 5: Dissociation of transferrin from the receptor at the PM (pH-regulated), modeled in COPASI (Table I).

For this experimental setup, all transferrin receptors at the PM (a total of 2,052 molecules) were initially loaded with Tf.Fe (all R switched to R-Tf.Fe). The internalization rate was calibrated to match experimental values [1]. Experimentally, radiolabeled transferrin was used to follow the transport of this protein. Hence, to monitor total transferrin in the model, all molecular forms of the protein (R-Tf.Fe, R-Tf, Tf.Fe and Tf) were combined for analysis. Fig. 2A shows that the simulations perfectly reproduce the experimental values. Lodish's study employed an alternative experimental approach using $^{59}Fe$. In this scenario, labelled iron dissociates from transferrin intracellularly, with the majority translocated into the cytosol. A fraction of the label is released into the medium due to Tf.Fe dissociation from receptors at the PM. To compare with this experimental approach, all molecular forms of iron in the model were combined (R-Tf.Fe, Tf.Fe, and Fe). The model successfully reproduces these experimental findings as well (Fig. 2B). For these simulations, cytosolic Fe levels were calculated as a difference as the cytosol was not explicitly represented in the model.

The kinetics of each molecular species are shown in Figure 2C, with values normalized to their maximum abundance in the system (indicated in the figure panels). R-Tf.Fe disappears rapidly, as most complexes are internalized and release their iron. A small fraction dissociates at the PM, resulting in soluble Tf.Fe that diffuses into the medium. R-Tf, on the other hand, is generated in acidic early endosomes and recycles to the PM, with a fraction retained in SE and RE compartments. Free Tf emerges as the most abundant final metabolite, accumulating predominantly in the medium. Notably, radiolabeled protein in the medium originates from two sources: Tf.Fe that dissociates from PM receptors and Tf that dissociates from receptors at neutral pH after intracellular transport and iron release in acidic compartments. Free Fe is a transient species, primarily pumped out of acidic organelles, with minimal release into the medium.

Several insights emerge from analyzing the dynamic model. While the average behavior of transferrin molecules aligns with Lodish's description, the process is stochastic. During a single round of endocytosis, only a small fraction of transferrin molecules reach the RE. Most molecules are recycled back to the PM from EE or SE compartments. Some complexes recycle without releasing their iron and are re-internalized, while others dissociate at the PM without entering the cell. In conclusion, the highly dynamic nature of the endocytic pathway profoundly influences the trajectory and fate of internalized complexes. These findings highlight the complexity of intracellular trafficking, a scenario likely applicable to other endocytosed molecules as well.

## Following Continuous Uptake and Subsequent Chase of Transferrin

Transferrin is widely used in cell biology as a standard marker for recycling processes. Continuous uptake and subsequent chase experiments are commonly employed to study recycling pathways. To mimic these experimental protocols, the simulation was run with Tf.Fe present in the medium for 60 minutes. Subsequently, all medium molecules, including Tf.Fe, were removed, and the simulation was continued for an additional 60 minutes.

In the presence of Tf.Fe in the medium, the cycle of R-Tf.Fe internalization and R-Tf recycling delivered iron to the cell and Tf to the medium in a linear manner (Fig. 4A and 4B). According to Lodish, cells internalize transferrin at a rate of approximately 9,500 molecules per minute over extended periods of time. Consistent with this, our model—which represents 1/30th of a complete cellular plasma membrane—released Tf at a rate of 300 molecules per minute (Fig. 4A). Within the first 15 minutes, R-Tf reached steady-state levels in EE and SE compartments. However, approximately 60 minutes were required to fully load RE (Fig. 4B). Only a small fraction of free Fe (around 1%) was lost to the medium (Fig. 4B).

After the medium was washed to remove all soluble metabolites at the plasma membrane, R-Tf.Fe was partially released into the medium while the remaining molecules were internalized. Most R-Tf present in EE and SE compartments rapidly recycled to the surface, while a fraction continued its journey to RE. Recycling from RE back to the plasma membrane required significantly more time. After 60 min, most of the transferrin has been released into the medium.

## Following the Transferrin Path on Board a Transferrin Receptor

The previous simulations demonstrated that the endomembrane system, along with the key molecules and kinetic parameters involved in iron delivery, accurately reproduces a wide range of experimental observations. These simulations also revealed significant heterogeneity among individual membrane-bound compartments. However, the final outcomes of independent simulations were consistent because they represent the average behavior of hundreds of individual molecular trajectories.

In addition, the model provides a unique opportunity: the ability to track single trajectories of a transferrin receptor within the endomembrane system. To achieve this, during a continuous uptake of transferrin, a single transferrin receptor molecule was labelled. This molecule behaved like a regular receptor, but during fission events, instead of partitioning according to differential affinity for membrane domains, it was incorporated into one of the resulting vesicles. The receptor's affinity for specific domains was used to calculate the probability of its delivery to a particular compartments.

Using this approach, the trajectory of this single receptor was tracked as it navigated the endomembrane system—from its incorporation into an early endosome (EE) to its return to the plasma membrane and subsequent internalization. Every 0.1 minutes, the simulation recorded detailed properties of the vesicle containing the receptor, including its cargo of soluble and membrane-bound molecules. Over one hour of continuous transferrin uptake, sixteen in-and-out trajectories of the receptor were documented (Fig. 4A).

The fusion, fission, and maturation dynamics of endosomes exposed the receptor to various scenarios. For example:

1. **First Trajectory** (Fig. 4B, left): The receptor entered an EE that acidified rapidly, converting R-Tf.Fe into R-Tf. The organelle's changing area indicated a fission event, followed by fusion with another vesicle. Approximately 2 minutes later, the receptor-containing endosome fused back with the plasma membrane.
2. **Second Trajectory** (Fig. 6B, middle): The receptor entered an EE where pH decreased, releasing Fe from R-Tf.Fe. Part of the EE matured into a SE that is sorted out by means of a fission event, incorporating the receptor. The SE's pH gradually increased, and a small region of the SE matured into RE. Soon after, this RE fused with another EE, triggering a new round of acidification. The resulting hybrid endosome carried characteristics of EE, SE, and RE compartments. Eventually, the receptor was segregated into an SE that fused with the plasma membrane.
3. **Third Trajectory**: The receptor moved through an EE that underwent maturation and fission mediate sorting into a SE compartment. Multiple fission and fusion events maintained the receptor in a hybrid endosome with EE, SE and RE characteristics. Most of the other receptors in the compartment were bound to Tf (R-Tf), almost all Fe was released into the cytosol, and a fraction lost Tf to become empty receptors (R). Finally, during a fission event, the receptor is incorporated in a RE where it stayed for several minutes before the endosome fused with the PM

These complex changes in the characteristics of the carrier organelle and its cargoes are captured in **Video 1**.

## Conclusions

The endomembrane system defines an internal work separated from the cytoplasm. This interconnected system is neither continuous nor homogeneous. Membrane domain maturation facilitates directional transport, fusion enables the mixing of compatible compartments, and fission drives the sorting of membrane domains and their associated cargoes. While detailed molecular mechanisms have been proposed for each of these processes, our understanding of how they integrate to support efficient intracellular transport remains incomplete. Mathematical models have suggested that the interplay of these three mechanisms is essential for various transport scenarios [11,12]. However, the absence of a consensus-based minimal model that adequately represents the dynamic nature of intracellular trafficking limits our understanding of key cellular processes within the endomembrane system. To make evident this gap, we extended our simulation to encompass a significant section of a cell and applied it to the transferrin pathway. The well-characterized intracellular transport of transferrin revealed an unexpected level of complexity when its molecular interactions were simulated within the context of the dynamic endocytic pathway.

Mathematical and computational models of molecular and cellular systems are invaluable tools for advancing our understanding of biological processes. Both all-atom and coarse-grained molecular dynamics simulations, for example, provide simplified representations of molecular interactions that replicate the behavior of complex molecular assemblies—assemblies that are often inaccessible to experimental methods or more detailed theoretical models. Similarly, studying cellular processes within dynamic organelles that operate over extended timescales requires a balanced approach, integrating holistic transport models with molecular details specific to particular phenomena. The Agent-Based Model (ABM) employed to simulate transferrin trafficking is particularly well-suited for such applications. Furthermore, this model can be expanded to represent a complete three-dimensional cell, incorporating diverse organelles within a dynamic cytoplasm, while allowing the inclusion of molecular interactions—all without exceeding the computational capacity of standard hardware.

Finally, we emphasize the critical importance of quantitative measurements of cellular properties and processes, such as those used to adjust the present model and published decades ago. These measurements are indispensable for refining models that aim to accurately represent cellular systems and their dynamic behaviors.

ACKNOWLEDGMENTS
FN and LSM were supported by CONICET (Consejo Nacional de Investigaciones Científicas y Técnicas, Argentina).

CONFLICT OF INTEREST
The author declare no potential conflict of interest.

Table I. Reactions implemented in COPASI

| Name | Reaction | Rate Law | reference |
| --- | --- | --- | --- |
| free Tf.Fe | R + Tf.Fe → R-Tf.Fe | Mass action | [1] |
| release Fe | R-Tf.Fe → R-Tf + Fe | pH-1s-Logistic[a] | [5] |
| free Tf | R-Tf → R + Tf | pH-1s-Logistic | [1] |
| Fe Transport | Fe endosome → Fe cytosol | pH-1s-Logistic | [8] |
| ProtonPump LE | → $H^+$ | Constant flux | * |
| ProtonPump EE | → $H^+$ | Constant flux | * |
| $H^+$ Leak | $H^+$ endosome ←→ $H^+$ cytosol | Mass action | * |

[a] logistic rate low to regulate the kinetics of the reaction according to the pH [9]
* Arbitrary chosen to get acidification kinetics as reported in [7] for DMT1 containing endosomes

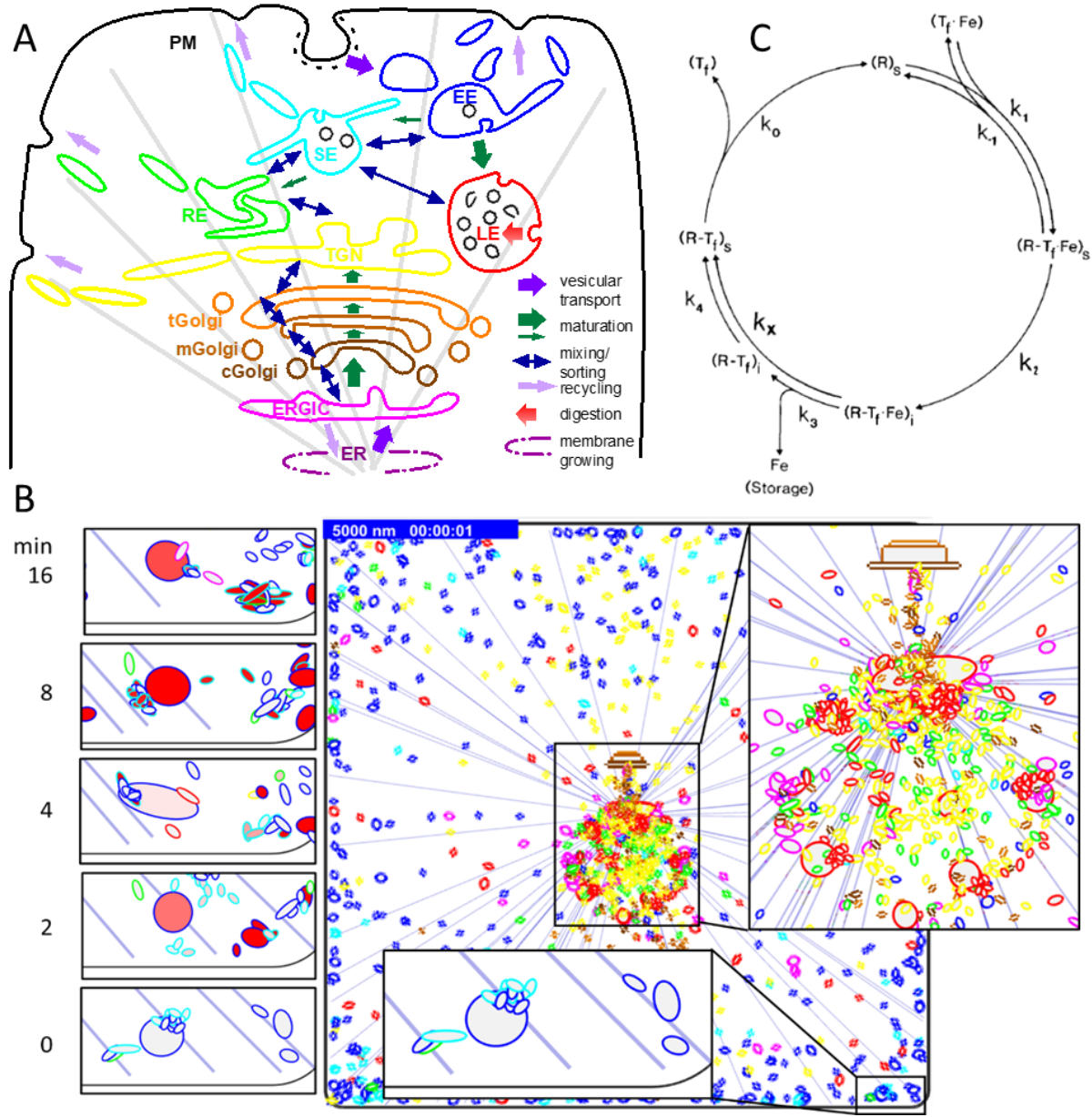

**Figure 1. Brief description of the model A)** The cartoon illustrates the compartments included in the model and their interactions: TGN (trans-Golgi network), tGolgi (trans-Golgi), mGolgi (medial Golgi), cGolgi (cis-Golgi), ERGIC (ER-Golgi intermediate compartment), ER (endoplasmic reticulum), EE (early endosomes), SE (sorting endosomes), RE (recycling endosomes), LE (late endosomes/lysosomes), and PM (plasma membrane). Dark violet arrows indicate vesicle budding from the ER and PM, while green arrows represent maturation from one membrane domain to the next. Dark blue bidirectional arrows represent the probability of fusion between compartments, with heterotypic fusion facilitating the mixing of compartments, which are subsequently sorted into distinct structures during vesicle or tubule budding. For simplicity, homotypic fusion (highest probability between organelles carrying the same membrane domain) is not depicted. Fusion of EE, RE, and TGN organelles with the PM, as well as ERGIC structures

with the ER, is shown with light violet arrows. The red arrow indicates the digestion of membranes and soluble content in the LE. **B)** Snapshot of a simulation showing organelles corresponding to all the compartments described in (A). Snapshot of a simulation showing organelles corresponding to all the compartments described in (A). Two regions are enlarged to highlight details. Changes in the organelles at the bottom right of the simulation are shown during the continuous uptake of transferrin (red content). The edges of each organelle are color-coded according to the most abundant membrane domain present. Straight blue lines represent microtubules, which orient the movement of organelles. **C)** Reproduction of Fig. 1 of the Lodish's original paper [1].

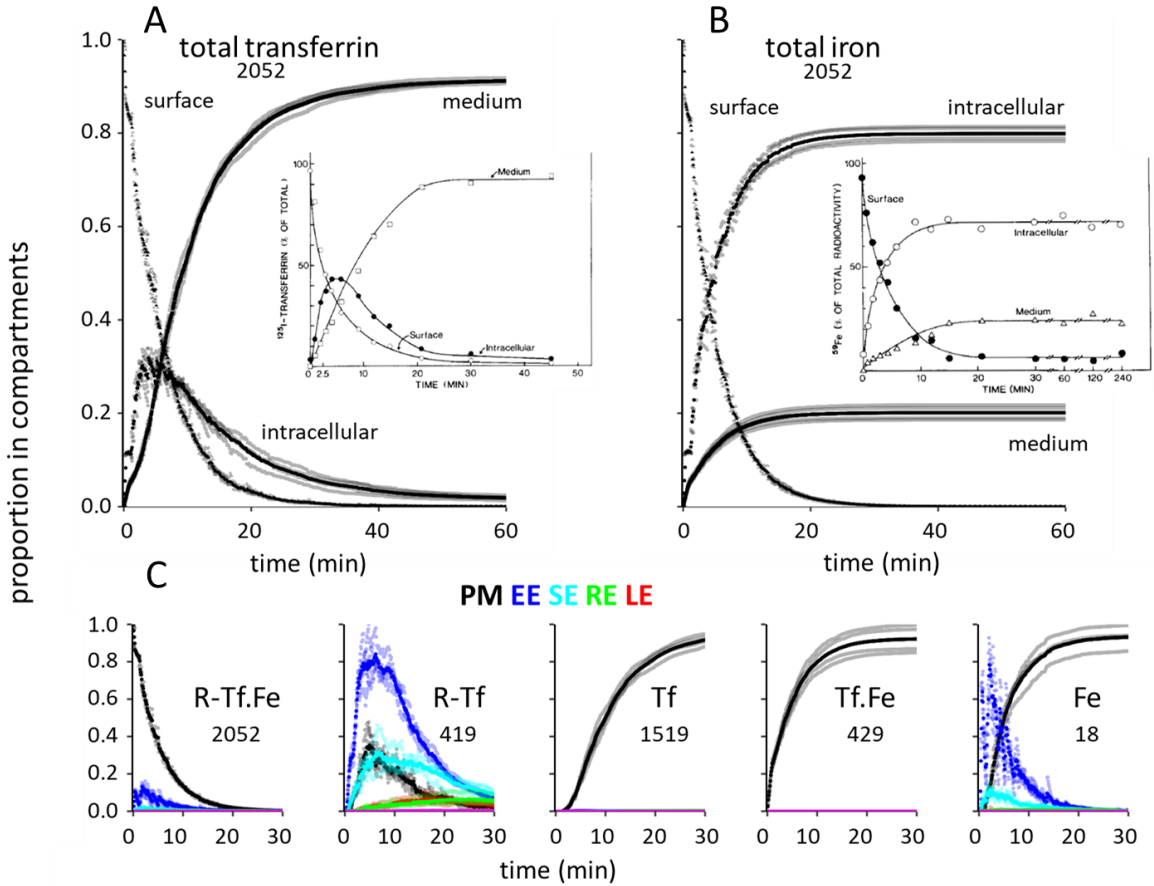

**Figure 2. Single cycle of transferrin internalization.** At the start of the simulations, receptors at the PM (2052 molecules) were loaded with transferrin (R-Tf.Fe). The model was run three times over 60 minutes. Individual simulations are shown in pale colors, with their average represented in bright colors. Values for each species (individual or grouped) are shown as proportions within each compartment, normalized to the maximum indicated in each panel. **(A)** Distribution simulating radiolabeled transferrin uptake, which combines R-Tf.Fe, R-Tf, Tf, and Tf.Fe. **(B)** Distribution simulating radiolabeled iron (59Fe) uptake, combining R-Tf.Fe, Tf.Fe, and Fe. Iron transported into the cytosol was inferred from the decay of the total iron within the system. Insets in panels (A) and (B) are slightly modified reproductions of Figures 2 and 3 from Lodish's original paper [1]. **(C)** Distribution of individual species across compartments (color-coded) during the simulations. In the PM, soluble species are released into a large-volume compartment (the medium). Values are normalized to the maximum number of molecules for each species, as indicated in each panel. Note that some species are only transiently present in the model, and Fe is rapidly transported out of the compartments.

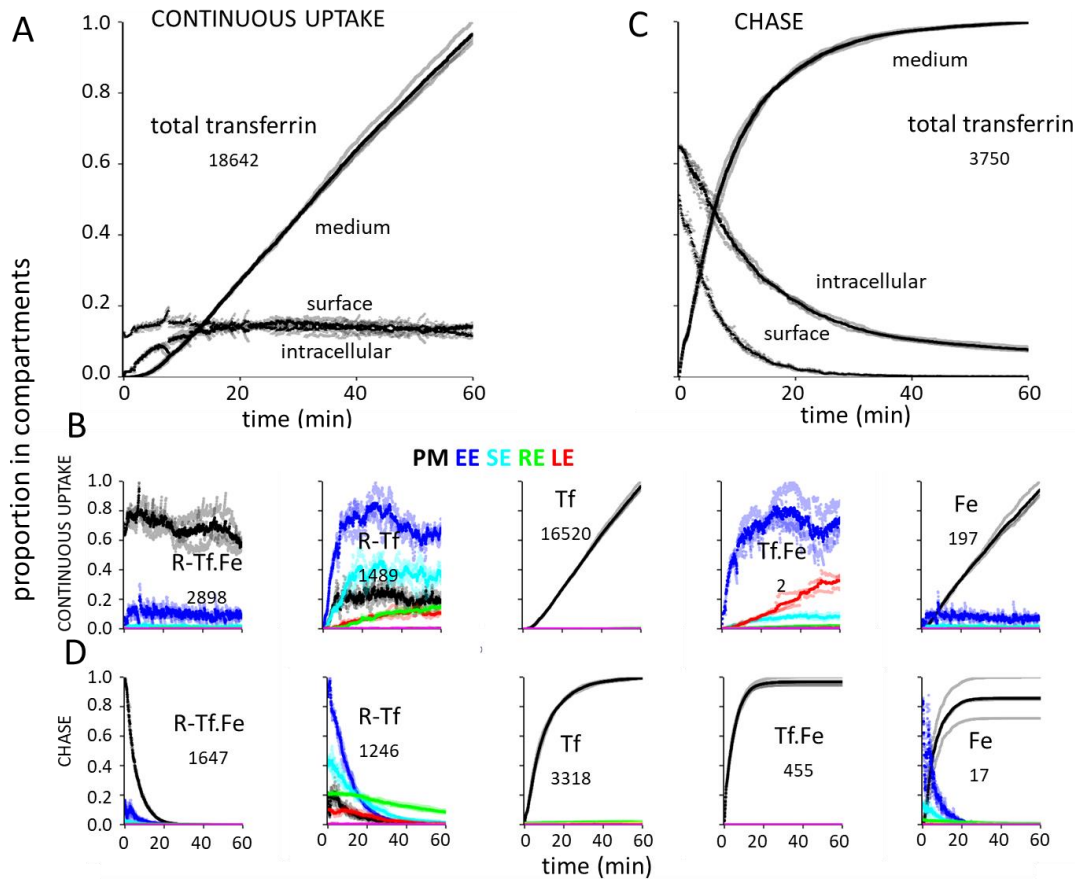

**Figure 3. Continuous uptake and chase of transferrin. (A, B)** Simulations were performed with saturating concentrations of Tf.Fe in the medium. Under these conditions, receptors at the PM incorporated the ligand whenever present (R → R-Tf.Fe). The model was run three times for 60 minutes. Individual simulations are shown in pale colors, with their averages represented in bright colors. Values for each species (individual or grouped) are presented as proportions within each compartment, normalized to the maximum indicated in each panel. **(A)** Distribution simulating radiolabeled transferrin uptake, combining R-Tf.Fe, R-Tf, Tf, and Tf.Fe.. **(B)** Distribution of individual species across compartments (color-coded) during the simulations. Only cell-associated Tf.Fe is plotted. **(C, D)** Following the 60-minute continuous uptake phase, all species in the medium were washed out, and the simulation continued for an additional 60 minutes. Individual simulations are shown in pale colors, and their averages in bright colors. Values for each species (individual or grouped) are presented as proportions within each compartment, normalized to the maximum indicated in each panel. **(C)** Distribution simulating radiolabeled transferrin uptake, combining R-Tf.Fe, R-Tf, Tf, and Tf.Fe. **(D)** Distribution of individual species across compartments (color-coded) during the chase period.

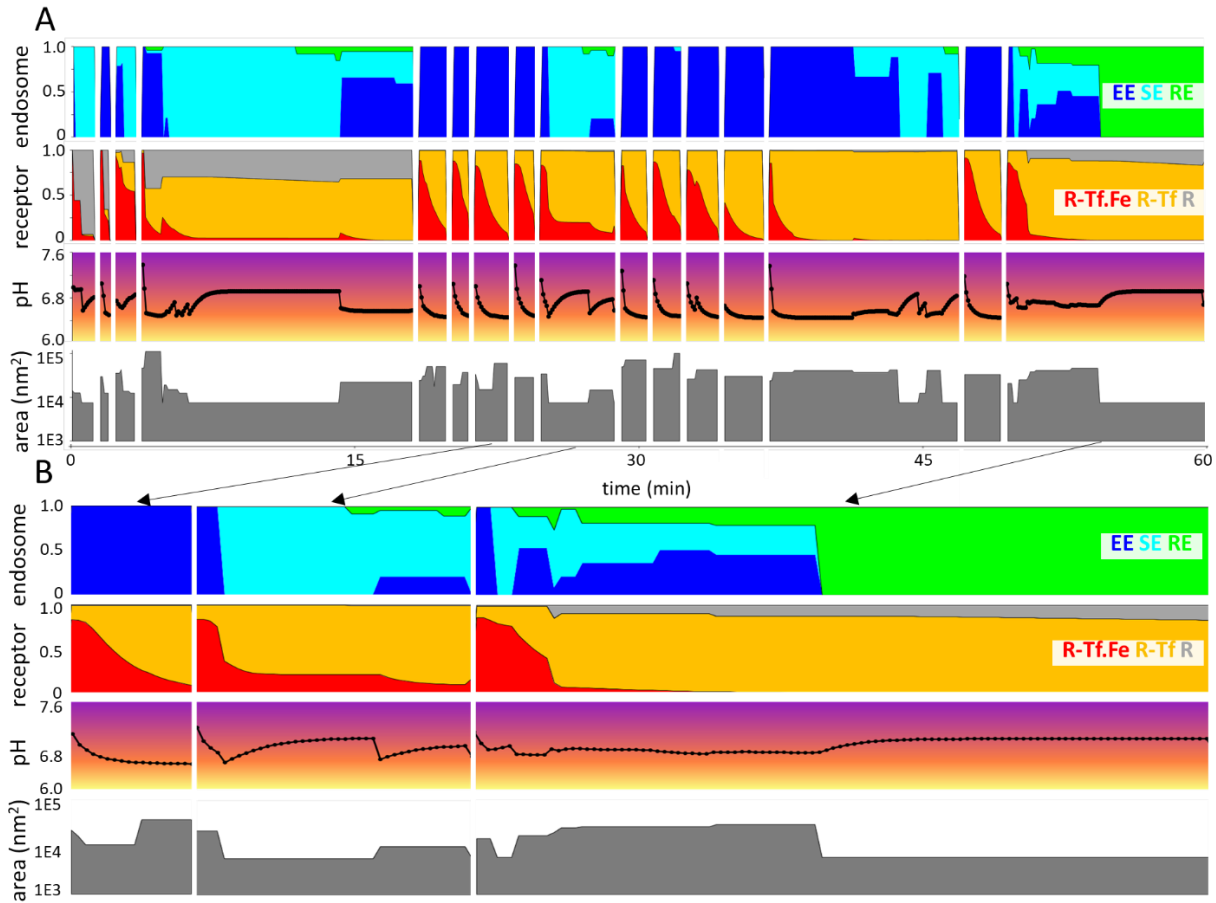

**Figure 4. Tracking the intracellular journey of a single transferrin receptor.**
(**A**) During a continuous uptake experiment, a single transferrin receptor molecule was labeled. The characteristics of the organelle carrying the receptor were recorded every 0.1 minutes. The plotted data includes: (**Endosome**) Proportion of the endosomal membrane associated with EE, SE, and RE. (**Receptor**) Proportion of the receptor in different states: unbound (R), bound to apotransferrin (R-Tf), or bound to iron-loaded transferrin (R-Tf.Fe). (**pH**) pH calculated from the number of $H^+$ molecules and the organelle volume. (**Area**) Area of the organelle in nm². Vertical white lines indicate the receptor's return to the PM and its incorporation into a newly formed EE. Note that shortly after internalization, the pH is neutral and most receptors are bound to iron-loaded transferrin (R-Tf.Fe). (**B**) Detailed view of three receptor trajectories inside the cell, highlighting changes in the parameters over time. See the text for a description of these trajectories.

**Video caption:** Graphical representation of changes in an organelle carrying a labelled transferrin receptor along the trajectories shown in Fig. 4A, starting at 19 minutes. The ellipse borders are colored according to the endosomal membrane domain, and the ellipse's size reflects the organelle's area and volume. The fill color indicates pH changes. Transferrin receptors are depicted as empty, bound to apotransferrin, or bound to iron-loaded transferrin.

.